\documentstyle[12pt,a4]{article}
\def\beq{\begin{equation}}
\def\eeq{\end{equation}}
\def\beqn{\begin{displaymath}}
\def\eeqn{\end{displaymath}}
\def\beqa{\begin{eqnarray}}
\def\eeqa{\end{eqnarray}}
\def\bega{\begin{array}}
\def\enda{\end{array}}

\def\fun#1#2{\lower3.6pt\vbox{\baselineskip0pt\lineskip.9pt\ialign
{$\mathsurround=0pt#1\hfil##\hfil$\crcr#2\crcr\sim\crcr}}}

\let\a=\alpha

                    \let\D=\Delta
\let\e=\epsilon

\let\theta=\theta

\def\ssty{\scriptstyle}
\def\sssty{\scriptscriptstyle}

\def\frac#1#2{{#1\over#2}}
\def\frc#1#2{\relax\ifmmode{\textstyle{#1\over#2}} \else$#1\over#2$\fi}
\def\frcc#1#2{\relax\ifmmode{\sssty{#1\over#2}} \else$#1\over#2$\fi}
\def\rlx{\relax\leavevmode}
\def\inbar{\vrule height1.5ex width.4pt depth0pt}
\def\sinbar{\vrule height1ex width.35pt depth0pt}
\def\ssinbar{\vrule height.7ex width.3pt depth0pt}
\font\cmss=cmssi12
\font\cmsss=cmssi12 at 7pt
\def\ZZ{\rlx\leavevmode
             \ifmmode\mathchoice
                    {\hbox{\cmss Z\kern-.4em Z}}
                    {\hbox{\cmss Z\kern-.4em Z}}
                    {\lower.9pt\hbox{\cmsss Z\kern-.36em Z}}
                    {\lower1.2pt\hbox{\cmsss Z\kern-.36em Z}}
               \else{\cmss Z\kern-.4em Z}\fi}
\def\Ik{\rlx{\rm I\kern-.18em k}}
\def\IC{\rlx\leavevmode
             \ifmmode\mathchoice
                    {\hbox{\kern.33em\inbar\kern-.3em{\rm C}}}
                    {\hbox{\kern.33em\inbar\kern-.3em{\rm C}}}
                    {\hbox{\kern.28em\sinbar\kern-.25em{\rm C}}}
                    {\hbox{\kern.25em\ssinbar\kern-.22em{\rm C}}}
             \else{\hbox{\kern.3em\inbar\kern-.3em{\rm C}}}\fi}
\def\IP{\rlx{\rm I\kern-.18em P}}
\def\ZZ{\rlx{\rm Z\kern-.25em Z}}
\def\IR{\rlx{\rm I\kern-.18em R}}
\def\Ident{\rlx{\rm 1\kern-2.7pt l}}
\def\ID{\relax{\rm I\kern-.18em D}}
\def\IF{\relax{\rm I\kern-.18em F}}
\def\IH{\relax{\rm I\kern-.18em H}}
\def\II{\relax{\rm I\kern-.17em I}}
\def\IN{\relax{\rm I\kern-.18em N}}
\def\IQ{\relax\,\hbox{$\inbar\kern-.3em{\rm Q}$}}

\def\gsim{\relax\leavevmode
     \ifmmode\mathchoice
        {\raise1pt\hbox{$>$} \kern-0.65em \lower3.5pt\hbox{$\ssty \sim$}}
        {\raise1pt\hbox{$>$} \kern-0.65em \lower3.5pt\hbox{$\ssty \sim$}}
        {\raise0.75pt\hbox{$\ssty >$} \kern-0.56em
                \lower3.5pt\hbox{$\sssty \sim$}}
        {\raise0.85pt\hbox{$\sssty >$} \kern-0.50em
                \lower2.7pt\hbox{$\sssty \sim$}}
     \else
           {$\raise1pt\hbox{$>$} \kern-0.65em
                \lower3.5pt\hbox{$\ssty \sim$}$}\fi}
\def\lsim{\relax\leavevmode
     \ifmmode\mathchoice
        {\raise1pt\hbox{$<$} \kern-0.65em \lower3.5pt\hbox{$\ssty \sim$}}
        {\raise1pt\hbox{$<$} \kern-0.65em \lower3.5pt\hbox{$\ssty \sim$}}
        {\raise0.75pt\hbox{$\ssty <$} \kern-0.56em
                \lower3.5pt\hbox{$\sssty \sim$}}
        {\raise0.85pt\hbox{$\sssty <$} \kern-0.50em
                \lower2.7pt\hbox{$\sssty \sim$}}
     \else
            {$\raise1pt\hbox{$<$} \kern-0.65em
                \lower3.5pt\hbox{$\ssty \sim$}$}\fi}

\begin{document}

{

\flushright{hep-ph/9510395}

}
\begin{center}
{\Large {\bf Strings with axionic content and baryogenesis \\}} \bigskip
\medskip
{\large Roberto Fiore} \\ {\it Dipartimento di Fisica, Universit\`a della
Calabria and\\ Istituto Nazionale di Fisica Nucleare, Gruppo collegato di
Cosenza\\ I-87030 Arcavacata di Rende (Cosenza), Italy }\\ \bigskip
{\large Luis Masperi and Ariel M\'egevand}\\ {\it Centro At\'omico Bariloche
and Instituto Balseiro \\ Comisi\'on Nacional de Energ\'\i a At\'omica and
Universidad Nacional de Cuyo \\ 8400 S. C. de Bariloche, Argentina}
\end{center}

\begin{abstract}
We describe different electroweak strings with axionic content, including
non-topological configurations calculated numerically, and show their
possible influence on baryogenesis indicating that they may constitute a
mechanism competitive to that of bubble nucleation with two Higgs-doublets.
\end{abstract}

\pagebreak

\section{Introduction}

There is much interest on the possibility that baryogenesis in the
universe had occurred during the electroweak (EW) phase transition \cite{ckn}
, even though particular GUT models with $B-L$ violation are not excluded
\cite{bf}. Due to the weakly first-order feature of this transition \cite
{dlhll} it may be important to have contributions from strings \cite{bd-bdt}
in addition or as alternative to the one coming from the bubble nucleation.
It seems also necessary to find sources of $CP$ violation stronger than
that of minimal Standard Model (SM).

In the present work we estimate the influence on baryogenesis of electroweak
strings which include axions. The relevance of this particle, possible
candidate for the dark matter, is twofold. On one side it may increase the
stability of the otherwise unstable electroweak strings. On the other hand
the $CP$ violation caused by the non-vanishing axion field in the phase
where baryogenesis is produced may be an alternative to the two
Higgs-doublets model or the supersymmetric extensions \cite{hn}. We analyze
the cases of the electroweak string which attracts axions inside its core
because of the smaller mass of these particles in the high temperature phase
\cite{ms}, the global axionic string formed at a higher scale which with the
addition of electroweak components assures the flux constancy of the
$Z$-magnetic field \cite{ds} \cite{mm}, and the non-topological
configuration \cite{fgmm} due to the interaction of the axion with
the neutral gauge boson $Z$. The existence of these last strings is shown
numerically in the thin-wall approximation, and it is found that in the case
in which the EW transition is of first order they become unstable below the
critical temperature  competing with bubbles to transform the metastable
symmetric phase into the broken symmetry stable one.

In section 2 we describe baryogenesis in the EW transition with the
possible contribution of general strings. In section 3 the strings with
axionic content are considered. The existence and
relevance of non-topological strings are numerically shown in section 4 and
brief conclusions are included in section 5.

\section{Baryogenesis at the electroweak transition and strings.}

The rate of generation of baryonic number per unit volume $n_B$ is \cite
{ckn}
\begin{equation}
\label{z1}{dn_B\over dt}\simeq -{{\frac{\Gamma}T}}\mu _B~,
\end{equation}
where $\Gamma $ is the probability per unit volume and time of anomalous
events, T is the temperature and $\mu _B$ is the chemical potential of
baryonic number, which is related to the densities of all left-handed
fermions.

Whereas the unsuppressed anomalous events rate in the symmetric phase is
\begin{equation}
\label{z2}\Gamma \sim \left( \alpha_WT\right) ^4~,
\end{equation}
with $\alpha _W$ electroweak coupling, $\mu _B$ depends on the possibility
that baryogenesis is local or that the bias produced by the $CP$
violation arrives by diffusion to a larger region according to details of
the transition. In the former case, if the $CP$ violation is given by the
relative phase $\theta $ between two Higgs doublets, the baryon number
density ideally produced by its change through the wall of an expanding
broken symmetry region which covers the whole space is \cite{b2}
\begin{equation} \label{z3}n_B\sim -{\frac \Gamma T}\left(
{{\frac{m_t}T}}\right) ^2\Delta \theta ~,
\end{equation}
where the top quark
is the most relevant of the species because of its high mass which gives
larger coupling.  In the latter case an effective $1/\alpha _W^2$ factor
appears \cite{bdpt} and the lepton $\tau $ becomes competitive to the top
quark because of its larger diffusion property.

Being the entropy density
\begin{equation}
\label{z4}s\sim g_{*}T^3~,
\end{equation}
with $g_{*}$ number of massless particle modes at the EW transition, we may
express the known present ratio of baryon to photon densities
$\eta $ as \begin{equation} \label{z5} \frac{\eta}7 = \frac{n_B}s \sim
\frac{\alpha_W^n}{g_*}\theta _{CP}\frac{V_{BG}}V~.
\end{equation} In Eq.(\ref{z5}) $n=4\ (2)$ for local
(non local) baryogenesis and $\theta _{CP}$ is the effective parameter,
including dynamical features, due to the change of the $CP$-violating angle
through the wall of the defect responsible for the universe non-equilibrium
period necessary to produce the matter-antimatter asymmetry. The suppression
factor $SF=V_{BG}/V$ is the ratio of the volume active for baryogenesis to
the total one.

If the EW phase transition is clearly of first order, its traditional
mechanism corresponds to the nucleation of bubbles of the broken-symmetry
phase in the metastable sea of the symmetric phase. In this case
baryogenesis is produced in the non-equilibrium region outside the wall of
the expanding bubble so that almost every point in space becomes active and
$SF\simeq 1$. Since $\alpha_W^2\sim 10^{-3}$ and $g_{*}\sim 10^2$, to
obtain from Eq.(\ref {z5}) the experimental result $n_B/s\sim 10^{-10}$ a
relevant contribution of $\theta _{CP} \sim 10^{-2}-10^{-5}$ is necessary.
This cannot come from the Kobayashi-Maskawa phase which gives \cite{ckn}
$\theta _{CP}\sim 10^{-20}$. But the spontaneous CP breaking due to the
two Higgs-doublets extension of the Standard Model allows easily the desired
result especially if non-local baryogenesis is envisaged.

Unfortunately for not too low Higgs mass the EW transition is perturbatively
calculated to be weakly of first order \cite{dlhll} or even of second order
with the minimal SM and the bubble mechanism might not be straightforward
 \cite{gk-gg} even though non perturbative effects seem to maintain the
discontinuous feature of the transition \cite{xxx} which is also enhanced
with the two Higgs-doublets models. It is in any case useful to think about
the contribution of cosmic strings \cite{bd-bdt}. If we consider strings
formed at a scale $\Lambda$ where a symmetry is broken maintaining the EW
one down to the lower scale $v$, its length and average separation from
one another will be the correlation length $\xi $ according to the Kibble
mechanism \cite{k}. After the EW transition inside its width $ \delta \sim
1/\sqrt{\lambda }v$, where $\lambda$ is the Higgs potential coupling
constant, there will be symmetric phase and the outer region will correspond
to the spontaneously broken vacuum so that, if the strings collapse to
decrease their energy, the active volume will be
\begin{equation} \label{z6}
V_{BG}\sim \delta ^2\xi \frac{V}{\xi ^3}~.
\end{equation}
The suppression
factor to be inserted into Eq.(\ref{z5}) would not seem too severe, but one
has to take into account that $\xi $ may be estimated as coming from the
correlation length at the formation time $t_f$ of a transition different
from the EW one. The length that one must introduce into Eq.(\ref{z6})
corresponds on the other hand to the EW time $t_{EW}$. If the string
dynamics is still in the friction stage due to the dominance of the universe
expansion term, the ratio will be \cite{k-ve}
\begin{equation} \label{z7}
\frac{\xi \left( t_{EW}\right) }{\xi \left( t_f \right) } \sim \left(
{t_{EW}\over t_f}\right) ^{5/4}~,
\end{equation}
with $\xi \left( t_f\right) \sim 1/\lambda ^{\prime }\Lambda$ and
$\lambda ^{\prime }$  the Higgs potential coupling constant responsible
for the string formation. Using the scale expansion for the radiation
epoch
\begin{equation}
\label{z8}
\frac{\Lambda}v \sim \left( {t_{EW}\over
t_f}\right) ^{1/2}~,
\end{equation}
the suppression factor
\begin{equation}
\label{z9}
SF=\left( \frac{\delta}{\xi \left( t_{EW}\right) } \right) ^2\sim {\lambda
^{^{\prime }2}\over \lambda} \left( {v \over \Lambda} \right) ^3
\end{equation}
is therefore very small unless $\Lambda $ is close to $v$.

\section{Strings with axionic content}

One may consider then $Z$-strings produced at the EW transition
\cite{v} which would seem to give a relevant baryogenesis. But these strings
are unstable for not too low Higgs mass \cite{jpv-ko} so that their
contribution to Eq.(\ref{z5}) would be in fact negligible unless new
ingredients succeed in stabilizing them.

One possibility is the inclusion of the axion, hypothetical candidate
for dark matter proposed as a dynamical field solution of the strong $
CP$ problem \cite {YYY}, which couples to gauge fields in a $CP$ violating
way such that $\theta _{CP}$ may be taken as coming from $\Delta a/f_{PQ}$,
where $\Delta a$ is the change of the axion field across the defect wall and
$f_{PQ}\simeq 10^{12}GeV $ is the scale at which the Peccei-Quinn symmetry
was broken \cite {r}. Axions have been already considered \cite{m} thinking
that for $T>1GeV$ they are massless so that its field may fluctuate and give
an interference with sphalerons which produces however a very small $\theta
_{CP}\sim 10^{-20}$.

We instead consider the possibility that the coherent change of the axion
field given by a global string may be the seed of formation of electroweak
strings. We remind that a global axionic string is a configuration where
along the $z$-axis there is a filament of unbroken Peccei-Quinn phase
defined by a chiral field $\Psi =0$ whereas out of it in the $x$-$y$ plane
$\Psi =f_{PQ}\exp \left( ia/f_{PQ}\right) $ with a dependence
$a/f_{PQ}=\alpha $ on the angle $ \alpha $ around the $z$-axis. When
temperature goes below the critical EW one, two kinds of strings may be
formed. One possibility is that a region in the $x$-$y$ plane, at a distance
$r$  from the axis of the global string where $ \Psi =f_{PQ}\exp (i\alpha )$
and the Higgs field $\varphi =0$, takes the form of a torus surrounded by
the broken EW phase $\left| \varphi \right| =\left| v\right|/ \sqrt{2}$ and
$\Psi =f_{PQ}$ because a small axion mass forces $a$ to the potential
minimum $a=0$. An alternative is that the $z$-axis filament of unbroken
Peccei-Quinn phase $\Psi =0$ and $\varphi =0$ enhances to a finite width
outside which, if there are two Higgs-doublets, the angle $a/f_{PQ}$
compensates the difference of the phases of $\varphi _1$ and $\varphi _2$
giving a topological reason of stability to the EW string \cite{ds}.

Apart from these possibilities, when the temperature approaches the critical
value from above, a torus in the $x$-$y$ plane may appear inside which the
EW broken phase $\varphi =v/\sqrt{2}$ is formed and together with $\Psi
=f_{PQ}\exp \left( i\alpha \right) $ simulates a Chern-Simons model through
its coupling with gauge fields \cite{fgmm} which stabilizes the resulting
string, whereas outside it we have the symmetric phase $\varphi =0$ and
$\Psi =f_{PQ}\exp \left( ia_0/f_{PQ}\right) $ with the approximation of an
average value $a_0$ for the axion field since the axion mass is practically
zero there.

We will consider these three strings in the above order. The fact that the
axion must be lighter in the higher temperature phase may be simulated by an
effective potential \cite{ms} \cite{mm}
of interaction of the Higgs $\varphi$ and axion $a$ fields
\begin{equation}
\label{z10}
U=\lambda \left( \varphi ^{\dagger }\varphi -\frac{v^2}{2}
\right) ^2+\left[ m_a^2f_{PQ}^2+\kappa \left( \varphi ^{\dagger }\varphi -
\frac{v^2}{2}\right) \right] \left( 1-\cos {a\over f_{PQ}}\right)~,
\end{equation}
where $m_a$ is the axion mass in the phase of EW broken symmetry and
$\kappa $ fixes its extremely small mass $m_a^{\prime }$ in the symmetric
phase according to
\begin{equation}
\label{z11}
m_a^{^{\prime }2}={\kappa \over 2}{v^2\over f_{PQ}^2}-m_a^2~.
\end{equation}
The only important features of
the effective potential $U$ are the phase nature of the axion field and the
fact that for the broken symmetry vacuum $ a=0$ whereas in the symmetric one
the axion field prefers a non-vanishing value, due to the fact that there
it needs not  compensate the argument of the determinant of the mass
matrix \cite{r}.

The first proposed string has magnetic flux of the neutral $Z$ field through
its core but it is unstable considering only EW fields for realistic
parameters of the minimal SM \cite{jpv-ko}. The increase of stability  due
to the addition of axions is twofold \cite{mm}. On one side the fact that,
analogously to bags, the disappearance of the string with the light axions
concentrated in its core would require the energy for building the higher
mass of axions in the broken symmetry phase.  On the other hand, and this is
of more relevance, the oscillation of the phase $a/f_{PQ}$ along the core
between $0$ and $2\pi$ gives a quasitopological reason for the stability of
these closed strings in analogy with the superconducting ones \cite{w}. It
is important to remark that whereas outside the string $|\varphi| =
v/\sqrt{2},\, a=0$, inside the core $\varphi \simeq 0,\,a/f_{PQ}\sim \pi $,
so that the variation of the $CP $ violating phase across the wall is
$\Delta \left( a/f_{PQ}\right) \sim 1$, i.e. large without need of two
Higgs-doublets.

Note that in the collapse of these strings the change $\Delta \theta $ has
the same sign as in the expansion of bubbles because in the bias for
baryogenesis the explicit $CP$ violation angle and the axion field  appear
with opposite signs. Regarding the suppression factor, it must be analyzed
more carefully when strings are formed close to the scale of a possibly
second order EW transition. In this case for the temperature $T_s$, smaller
than the critical one $T_c$, when the string can be identified, i.e., when
its length becomes larger than the Ginzburg length \cite{kv}, the width will
be
\begin{equation}
\label{z12}
\delta \sim {1 \over m_H\left( T_s\right) }=\frac 1{\lambda ^{4/7}}\left(
{m_P\over T_c}\right) ^{3/14}{1 \over T_c}~,
\end{equation}
where $m_P$ is the Planck mass, and the length is
\begin{equation}
\label{z13}
\xi \sim \frac{1}{\lambda ^{3/7}}\left( {m_P\over T_c}\right)
^{2/7}{1 \over T_c}~.
\end{equation}
Therefore the suppression factor turns out to be
\begin{equation}
\label{z14}
SF=\left( {\delta \over \xi} \right) ^2 \sim \frac{1}{\lambda ^{2/7}} \left(
{T_c\over m_P}\right) ^{1/7}\relax\leavevmode \ifmmode\mathchoice
{\raise1pt\hbox{$<$} \kern -0.65em \lower3.5pt\hbox{$\ssty \sim$}}
{\raise1pt\hbox{$<$} \kern-0.65em \lower3.5pt\hbox{$\ssty \sim$}}
{\raise0.75pt\hbox{$\ssty <$} \kern-0.56em \lower3.5pt\hbox{$\sssty \sim$}}
{\raise0.85pt\hbox{$\sssty <$} \kern-0.50em \lower2.7pt\hbox{$\sssty \sim$}}
\else                                      { $\raise1pt\hbox{$<$}
\kern-0.65em \lower3.5pt\hbox{$\ssty \sim$}$}\fi\; 10^{-2}
\end{equation}
for $T_c\sim v$. As a result $n_B/s$ would be of the correct order of
magnitude, if kinematical factors  are similar to those of bubbles so that
$\theta_{CP} \sim 10^{-3}$, and non-local effects
enhance baryogenesis.

For the formation of the string it is favourable that the axion is heavier
in the broken symmetry phase. This might occur in a more
definite way at a lower scale $\sim 1GeV$ where $QCD$ effects are
particularly relevant. We must note that the very strong true magnetic
fields present at the electroweak scale may cause the instability of the
broken symmetry vacuum \cite{ao-ao} delaying the phase transition to a lower
scale closer to the $QCD$ one where, due to the universe expansion, the
magnetic field is weaker. If this occurs we may trust more in our string,
paying however the price of a smaller $T_c$ in Eq.(\ref{z12}) with a slight
decrease of $SF$.

The second proposal for formation is one to one related to the
original axionic string, at variance from the previous case where many EW
strings may appear from the same global one. Therefore the suppression
factor (\ref {z9}) will be small due to the formation scale $f_{PQ}$ much
larger than the EW one $v$.

These strings base their stability on a topological argument which requires
two Higgs-doublets with the potential \cite{ds}
$$
V\left( \varphi _i,\Psi \right) =\sum_{i=1}^2\frac{\lambda _i}
4\left( \varphi _i^{\dagger }\varphi _i-v_i^2\right) ^2+\frac \lambda
2\left( \varphi _1^{\dagger }\varphi _1\right) \left( \varphi _2^{\dagger
}\varphi _2\right)
+\frac{\lambda ^{\prime }}2\left( \varphi _1^{\dagger
}\varphi _2\right) \left( \varphi _2^{\dagger }\varphi _1\right)
$$
\begin{equation}
\label{z14'}
+f_{PQ}v_1v_2-\frac 12\left[ \left( \varphi _1^{\dagger }\varphi _2\right)
\Psi +h.c.\right]
\end{equation}
which forces the axion field to compensate the difference of Higgs angles,
i.e. $\frac a{f_{PQ}}=\theta _1-\theta _2$, outside their core.

It must be noted that with the potential  (\ref{z14'}) there are not
$CP$ violating terms associated with Higgs doublets so that the bias for
baryogenesis is entirely due to the axion coupling with gauge fields where,
because of the global origin of the string, $\Delta a/f_{PQ}=2\pi$ around
its axis.  If $CP$ violating terms of the type $Re(\varphi _1^{\dagger
}\varphi _2)-v_1v_2\cos \left( \theta _1-\theta _2\right) $ and $Im(\varphi
_1^{\dagger }\varphi _2)-v_1v_2\sin \left( \theta _1-\theta _2\right) $ were
added to Eq.(\ref {z14'}), the strings would become unstable and the bias
due to $\theta _1-\theta _2$ would tend to compensate that caused by
the axion field.

The stability of the axionic strings with the potential (\ref{z14'})
should be assured until the $QCD$ scale, giving time to a delayed EW
breaking, when domain walls attach to them making maximum the decay by
radiation of axions \cite{bs-ve}. However, even though the dynamics of
global strings is difficult because of long range interactions, one may
expect a small contribution to baryogenesis due to the suppression factor.

The last string we will consider is the non-topological configuration,
regarding a single Higgs doublet, which is made possible by the interaction
of the axion with the gauge field $Z$. The existence of this configuration
without a Maxwell kinetic term was numerically proved in 2+1 dimensions
where the previous interaction was replaced by a Chern-Simons term
\cite{fgmm} and evidences were also given including the Maxwell term
\cite{b}. Adding the third spatial dimension, the role of the Chern-Simons
constant is played by the constant gradient of the axion field along the
$z$-axis inside the string, remnant of the axionic string which existed
when the configuration was formed. The numerical computation of this case is
described in the next section.

\section{Non-topological strings}

The energy of a static cylindrically symmetric
configuration including the coupling of the axion field only with the
gauge field $Z$ for small enough electric field turns out to be
\begin{equation} \label{z15}
E=\int d\vec
r\left[ \frac12\vec B_Z^2+\frac14\left( \frac{\alpha}{f_{PQ}g} \frac{\vec
\nabla a\cdot \vec B_Z}{\varphi} \right) ^2+\left( g\varphi Z_\theta \right)
^2+\left( \vec \nabla \varphi \right) ^2+\left( \vec \nabla a\right)
^2+U\right]
\end{equation}
where the potential $U$ is shown in Eq.(\ref{z10}) and we look
for the behaviour in the $x-y$ plane \begin{equation} \label{z16}
\begin{array}{lllll} \varphi \rightarrow 0 &  & a\rightarrow a_o & , &
r\rightarrow \infty~, \\ &  &  &  &  \\ \varphi \rightarrow \frac
v{\sqrt{2}} &  & a\propto z & , & r\rightarrow 0~.
\end{array} \end{equation}
The reason of the possible stability for $T>T_c$ is that, at variance from
the purely Maxwell kinetic term, the Chern-Simons-like contribution allows
$\vec B_Z\neq 0$ only in the region where $\varphi \neq 0$, i.e.  inside the
string core. Moreover, even though $U$ increases the energy for $a$
increasing monotonically along the string, for closed loops the change $
\Delta a=N2\pi f_{PQ}$ produces again a quasitopological stabilization.

Taking the thin wall approximation for the change of fields $a$ and
$\varphi$ in a common width $\varepsilon\ll R$, with $R$ radius of the
string, the equations of motion give the increase of magnetic field as an
expansion in powers of $r$
$$
B_Z(r)\simeq B_Z(0)\left( 1+\frac12 g^2v^2r^2\right)~.
$$
According to how many times the radius is larger than the width, the
magnetic field will tend to concentrate in the wall. We will take this
extreme case which is the most disfavourable for the stability of the
string and denote the flux as $2\pi h_0$. The minimization of the energy
described by Eq.({\ref{z15}) with respect to $R$, which  is expected to vary
with $z$, gives $$ \frac{2\pi}{\varepsilon}(v^2+a^2_0)+2\pi U_0 R-\left[
\frac{2{\pi}^3}{\varepsilon}\left(\frac{\alpha h_0}{gvL}\right)^2+
\frac{\pi\varepsilon}{8}(gvh_0)^2+\frac{\pi h^2_0}{\varepsilon}\right]
{1\over R^2} $$ $$
-\frac{2\pi}{\varepsilon}(v^2+a^2_0)\dot{R}^2-\frac{4\pi}{\varepsilon} (v^2+
a^2_0)R\ddot{R}-\frac{(4\pi)^2f_{PQ}a_0}{\varepsilon L}R\dot{R} $$
\begin{equation} \label{z17} -\frac{2 \pi h^2_0}{\varepsilon}{\ddot{R}\over
R}+\frac{\pi h^2_0}{\varepsilon} {\dot{R}^2\over R^2} = 0~,
\end{equation}
where $L$ is the string length, dots represent derivatives
with respect to the string axis $z$ and inside the core the potential is $$
U_0 = m^2_af^2_{PQ}\left(1-cos{2\pi\over L}z\right)~.$$ We are considering
$U=0$ outside the string to simulate the inclusion of temperature effects
such that $T=T_c$. Eq.(\ref{z17}) has been solved first for the terms
independent of $z$ and without $U_0$, together with the one which gives the
minimization of the energy with respect to the wall width $\varepsilon$,
valid if $L\gg 10^{-4}GeV^{-1}$, \begin{equation} \label{z18} \epsilon
^2=\frac{h_0^2+2a_0^2R_0^2}{\frac{\left( gvh_0\right) ^2}8+2\left( \frac{\pi
f_{PQ}}L\right) ^2R_0^2}, \end{equation} giving the constant radius $R_0$.
With the resulting $\varepsilon$ and taking $R_0$ as input we solved the
complete Eq.(\ref{z17}) with the conditions $R(0)=R(L)$ looking for
solutions such that $\dot R(0)=\dot R(L)$ in order to have periodic
conditions and consider the string as a closed loop. To allow this
interpretation we require $L \mathrel{\mathpalette\fun >} 10R$ and look for
solutions with only one oscillation of $a$ along the loop in correspondence
to lowest excited states.

In Table 1 we show the parameters of the found solutions with accepted
values of the physical constants appearing in Eq.(\ref{z17}). These
solutions are very close to the constant ones. Even though the flux is high,
it corresponds for many
of the indicated solutions to magnetic fields $|\vec B_Z| < 10^{24}G$ which
is the critical value \cite{ao-ao} for the destabilization of the broken
symmetry vacuum. The upper limit for $L$ corresponds to the fact that a
single oscillation $0\leq a/f_{PQ} \leq 2\pi $ does not allow to stabilize
too long strings. The lower limit for $L$ comes from the need that $\epsilon
< R$.

\vskip0.2cm

\begin{center}
\begin{tabular}{cccccccc}
\hline
$L$&$R/L$&$\e/L$&$h_0$\\
\hline
$10^{3}$   &   0.077  &  $0.41\times 10^{-1}$    &    $10^{13}$\\
$10^{5}$   &   0.077  &  $0.41\times 10^{-2}$    &    $10^{14}$\\
$10^{7}$   &   0.077  &  $0.41\times 10^{-3}$    &    $10^{15}$\\
$10^{9}$   &   0.077  &  $0.41\times 10^{-4}$    &    $10^{16}$\\
$10^{11}$  &   0.077  &  $0.41\times 10^{-5}$    &    $10^{17}$\\
$10^{13}$  &   0.077  &  $0.41\times 10^{-6}$    &    $10^{18}$\\
$10^{15}$  &   0.077  &  $0.41\times 10^{-7}$    &    $10^{19}$\\
$10^{17}$  &   0.077  &  $0.41\times 10^{-8}$    &    $10^{20}$\\
$10^{19}$  &   0.077  &  $0.41\times 10^{-9}$    &    $10^{21}$\\
 \hline
 \end{tabular}
\end{center}
\begin{center}
{\bf Tab.~1.} {\it For the string solutions of Eq.(\ref{z17}) we report the
length $L$ measured in $GeV^{-1}$; the ratio, with respect to $L$, of the
mean radius $R$, and the width $\e$ defined in Eq.(\ref{z18}), together with
the $Z$-magnetic flux $h_0$. The following constants have been used:
$\a=1/137$, $g=0.3$, $ v=250GeV$, $f_{PQ}=10^{12}GeV$, $a_0=10^{14}GeV$ and
$m_a=10^{-14}GeV$.}
\end{center}

At variance from bubbles, non-topological strings (NTS) are produced when $
T>T_c$. If the EW transition is of first order, when the system cools below $
T_c$ the non-topological string survives and becomes first metastable and
finally unstable expanding as the bubble.

It is important to compare the contribution of critical bubbles and NTS to
the phase transition through their Boltzmann probability factor $exp(-E/T)$.

For the former ones, their energy is infinite for $T=T_c$ and decreases for
$ T<T_c$. It is easy to estimate the bubble energy minimizing the sum of the
negative volume contribution and positive surface one. We get
\begin{equation}
\label{z19}
E_b = \frac{16}{3}\pi \frac{v^6}{\epsilon_b^3(\D U)^2}~,
\end{equation}
where $\D U$ is the difference of the energy density for symmetric and
broken symmetry vacua and $\epsilon_b \sim 100^{-1}GeV^{-1}$ is the bubble
wall width.

Regarding the string energy for $T<T_c$ its most important contributions
come from the negative volume term which must be included in $U$ of
Eq.(\ref{z15}), the surface and Maxwell terms. The equality of the bubble
and string energy occurs when
$$
\begin{array}{lll}
& \Delta U_{EQ}\simeq 10^{-4}GeV^4 & \mbox{~~for the shortest strings}
\end{array}
$$
and
$$
\begin{array}{lll}
& \Delta U_{EQ}\simeq 10^{-16}GeV^4 & \mbox{~~for the longest ones.}
\end{array}
$$

When $\D U < \D U_{EQ}$ one may expect that the NTS are relevant because
their energy is smaller than the bubble one. However one must verify that
the critical difference of vacua energy densities $\D U_c$, which makes the
NTS unstable, is smaller than $\D U_{EQ}$. One may estimate $\D U_c$ taking
the most relevant terms for the string energy and putting the condition that
the second derivative with respect to the mean radius is not positive in
order to avoid the local metastable minimum. Whereas for the shortest
strings $\D U_c \gg \D U_{EQ}$ indicating that they remain metastable while
bubbles expand and produce the phase transition, for the longest strings
$\Delta U_c \simeq 10^{-21}GeV^4\ll \Delta U_{EQ}$. Therefore the latter
become unstable very soon and contribute to the phase transition more than
bubbles in the initial stage. But one has to note that bubbles which equal
the energy of  the longest strings are extremely large and have $E_b \sim
10^{52}GeV$ i.e. of the order of Earth mass so that most of the phase
transition will presumably occur for lower temperatures when bubbles can be
easily formed by thermal fluctuations.

Our model for the non-topological string has been oversimplified
particularly in having fixed  a common wall width  for the Higgs and axion
fields giving therefore a very high contribution of kinetic energy of the
latter, and also in not taking into account in detail temperature effects on
the potential to describe the cosmological evolution. It is however
interesting to have shown that there is a period during which the longest
NTS expand and have energy smaller than the bubble one.

Since the NTS expand as the bubbles they have a suppression factor $SF\sim
1$. On the other hand the change through their wall of the axion field is on
the average $\Delta (a/f_{PQ})\sim \pi$ having the correct sign for the
baryogenesis bias. In this sense they might replace the two Higgs-doublets
except for the fact that the latter are probably needed to assure the first
order feature of the EW transition.

\section{Conclusions}

We have seen that electroweak strings originated in axionic ones increase
their stability and may contribute to the baryogenesis even if the
transition is of second order.

The addition of the axion to the standard  model represents an extension
smaller than the others proposed to explain the observed matter-antimatter
asymmetry.

The limitation to the usefulness of these strings, apart from the obvious
question regarding the existence of the axion, may be the possibility of
other decay channels which decrease their number as well as the necessity of
analyzing the details of the probability of their formation.

Through a careful study of the cosmological evolution of axionic strings it
may be possible to determine their quantitative influence on baryogenesis.

\vskip 1.5cm \underline {Acknowledgement}: One of us (L.M.) thanks the
Dipartimento di Fisica della Universit\`a della Calabria and the Istituto
Nazionale di Fisica Nucleare - Gruppo collegato di Cosenza for their warm
hospitality while part of this work was done. This research has been supported
in part by the Ministero italiano dell'Universit\`a e della Ricerca
Scientifica e Tecnologica and in part by CONICET, Argentina, of which A.M. is
fellow, through grant PID 3965/92.

\end{document}